\documentclass[prl,twocolumn,showpacs,amsmath,amssymb]{revtex4-1}

\usepackage{times}
\usepackage{graphicx}

\newcommand{\beq}{\begin{equation}}
\newcommand{\eeq}{\end{equation}}

\newcommand{\ba}{\begin{eqnarray}}
\newcommand{\ea}{\end{eqnarray}}
\newcommand{\msun}{M_\odot}

\def\mnras{Mon. Not. R. Astron. Soc.}
\def\aap{Astron. Astroph.}
\def\apjl{Astroph. J. Lett.}

\def\gs{\mathrel{\lower0.6ex\hbox{$\buildrel {\textstyle >}\over{\scriptstyle \sim}$}}}
\def\ls{\mathrel{\lower0.6ex\hbox{$\buildrel {\textstyle <}\over{\scriptstyle \sim}$}}}

\begin{document}

\title{Strong lensing of gravitational waves as seen by LISA}

\author{M. Sereno$^{1,2}$}
\email{mauro.sereno@polito.it}
\author{A. Sesana$^{3}$}
\author{A. Bleuler$^{4}$}
\author{Ph. Jetzer$^{4}$}
\author{M. Volonteri$^{5}$}
\author{M.~C. Begelman$^{6}$}
\affiliation{$^{1}$ Dipartimento di Fisica, Politecnico di Torino, Corso Duca degli Abruzzi 24, 10129 Torino, Italia}
\affiliation{$^{2}$ INFN, Sezione di Torino, Via Pietro Giuria 1, 10125, Torino, Italia}
\affiliation{$^{3}$Max Planck Institute for Gravitationalphysik (Albert Einstein Institute), Am M\"uhlenberg, 14476, Golm, Germany}
\affiliation{$^{4}$Institut f\"{u}r Theoretische Physik, Universit\"{a}t Z\"{u}rich, Winterthurerstrasse 190, 8057 Z\"{u}rich, Switzerland}
\affiliation{$^{5}$Department of Astronomy, University of Michigan, Ann Arbor, MI 48109, USA}
\affiliation{$^{6}$JILA, 440 UCB, University of Colorado at Boulder, Boulder, CO 80309-0440, USA}


\begin{abstract}
We discuss strong gravitational lensing of gravitational waves from merging of massive black hole binaries in the context of the LISA mission. 
Detection of multiple events would provide invaluable information on competing theories of gravity, evolution and formation of structures and, with complementary observations, constraints on $H_0$ and other cosmological parameters. Most of the optical depth for lensing is provided by intervening massive galactic halos, for which wave optics effects are negligible. Probabilities to observe multiple events are sizable for a broad range of formation histories. For the most optimistic models, up to $\ls 4$ multiple events with a signal to noise ratio $\gs 8$ are expected in a 5-year mission. Chances are significant even for conservative models with either light ($\ls 60\%$) or heavy ($\ls40\%$) seeds. Due to lensing amplification, some intrinsically too faint signals are brought over threshold ($\ls 2$ per year).
\end{abstract}


\maketitle

The space-based Laser Interferometer Space Antenna \citep[ LISA]{dan+al96} is expected to observe gravitational waves (GWs) and open a new window for astronomy. Coalescencing massive black hole binaries (MBHBs) with total masses in the range $10^3$-$10^7~M_\odot$ out to $z\sim 10$-$15$ are expected to provide the loudest GW signals at LISA frequencies, $f\sim \mathrm{mHz}$ \citep{lisa_sources}. In the standard cold dark matter (CDM) hierarchical cosmology, MBHBs form in large number during the multiple mergers experienced by their host galaxies. LISA should detect from few to several hundreds coalescences per year. Within such promising premises we discuss a potential new chapter for LISA science: multiple imaging of distant sources by intervening lensing galaxies. Gravitational lensing statistics, in either quasar or radio-galaxy surveys, is usually considered on a sample of several thousands of sources. Due to the unprecedented high redshift of LISA sources, and the related very high optical depth for lensing, multiple events are possible even for hundreds of detections. Astrophysical and theoretical rewards might be valuable. $i)$ Lensing of GWs would be another impressive confirmation of the theory of general relativity. How GWs propagate near a massive body might shed light on competing theories of gravity. $ii)$ Constraints on cosmological parameters might be obtained in the range $z\gs 10$. Measurements of time-delay, which can be accurately determined for transients \cite{statistics}, might be very useful. $iii)$ Lensing statistics might inform of the growth and structure of mass halos at $z\ls 3$. The expected number of events strongly depends on the form and evolution of the galaxy number density. $iv)$ The magnification effect could help in finding electromagnetic counterparts and observing objects otherwise too distant or too faint.
 
Strong lensing by ground-based GW detectors was discussed in \citep{wan+al96}. Here, we consider a lower frequency range where wave optics could play a role. Diffraction in lensing of GW is well understood \citep{gw_diff} and is effective only for small lenses \citep{eff_lenses}. Our study on strong lensing is complementary to those on the weak lensing distortion of GW signals, which mainly focused on their use as standard sirens \citep{sirens}. By default, we assume a flat $\Lambda$CDM model with $\Omega_\mathrm{M}=1-\Omega_\Lambda=0.3$, and $H_0=70~\mathrm{km~s}^{-1}\mathrm{Mpc}^{-1}$.

{\it Lenses.} Statistics of strong lenses is a well assessed astronomical tool \citep{statistics}. Here, we mainly follow \citep{ser05,zh+se08}. The differential probability of a source to be lensed is
\beq
\label{stat1}
\frac{d^2 \tau}{d z_\mathrm{d} d \sigma} = \frac{d n}{d \sigma}(\sigma, z_\mathrm{d}) s_\mathrm{cr}(\sigma, z_\mathrm{d})\frac{cd t}{d z_\mathrm{d}}(z_\mathrm{d}) ,
\eeq
where $\sigma$ and $z_\mathrm{d}$ are the velocity dispersion and the redshift of the deflector, respectively, $s_\mathrm{cr}$ is the cross section and $d n/d \sigma$ is the differential lens number density. $d n/d \sigma$ can be modeled as a modified Schechter function \citep{schechter}. This is accurate up to $z_\mathrm{d}\ls 1$, where galaxies provide the bulk of the cross section for LISA sources. We took a conservative approach. Together with a constant comoving number density, we also considered a pessimistic scenario, i.e., an evolving case with a smaller number of lenses at high $z$ \citep{cha07}.

We account only for lensing by early type galaxies, modelled as singular isothermal spheres (SISs). Their Einstein radius is $R_\mathrm{E}= 4\pi (\sigma/c)^2 D_\mathrm{d}D_\mathrm{ds}/{D_\mathrm{s}}$; $D_\mathrm{ds}$, $D_\mathrm{d}$ and $D_\mathrm{s}$ are the angular diameter distances between the deflector and the source, and the observer and the lens or the source, respectively. Two images form at $x_{\pm}=y\pm1$ if $y< 1$, with flux magnification $\mu_{\pm}=(1/y)\pm1$; $x$ and $y$ are the image and the source position normalized to the Einstein angular radius. The GW form is amplified by $A_{\pm}=\sqrt{\mu_{\pm}}$. $y$ can be measured through the ratio of the $A_{\pm}$. The delay between the arrival time of the images, $\Delta t=t_- -t_+$, is
\begin{equation}
\label{sis2}
\Delta t  =  \Delta t_z y , \ \ \ \ \Delta t_z  \equiv \frac{32 \pi^2}{c} \left( \frac{\sigma}{c}\right)^4 \frac{D_\mathrm{d} D_\mathrm{ds}}{D_\mathrm{s}} (1+z_\mathrm{d}).
\end{equation}
For $\sigma \simeq 200~\mathrm{km~s}^{-1}$, $z_\mathrm{d} \simeq 5$ and $z_\mathrm{s} \simeq 10$,  $\Delta t_z\simeq 100~\mathrm{days}$. LISA angular resolution is quite poor, $\sim 10'$, but lenses are expected to be very massive and luminous. In case the deflector is located, one can precisely constrain the position of the source and try to detect it with follow-up observations. Combining measurements of $\Delta t$, $z_\mathrm{d}$ and $z_\mathrm{s}$, and $D_\mathrm{s}$ from the GW analysis might allow us to constrain $H_0$ and dark energy.  A cosmographic approach independent of time-delay is based on lensing statistics \citep{statistics}.

\begin{figure}
         \scalebox{0.75}{\includegraphics{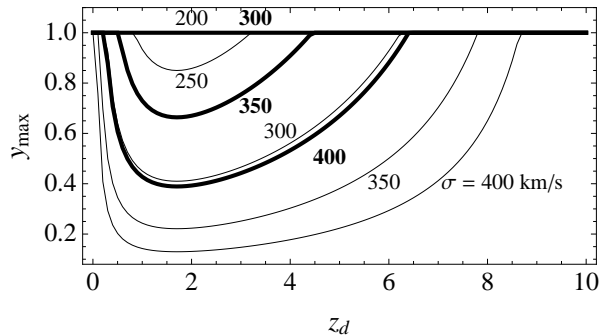}}
        \caption{$y_\mathrm{max}$ as a function of $z_\mathrm{d}$ for $z_\mathrm{s}=10$.   
        Thick and thin lines are for $t_{+}= 2$ or $4~\mathrm{years}$ from the survey kick-off.}
        \label{fig_1_y_max_from_delta_T}
\end{figure}

{\it The cross section} for lensing is $s_\mathrm{cr}=\pi R_\mathrm{E}^2 (y_\mathrm{max}^2-y_\mathrm{min}^2)$, where $y_\mathrm{min}<y<y_\mathrm{max}$ is the allowed range to form detectable multiple images. $y_\mathrm{max}$ depends on the lens mass and redshift, the threshold signal to noise ratio ($\mathrm{SNR_\mathrm{th}}$), the unlensed amplitude ($\mathrm{SNR_{int}}$), see Eq.~\ref{yMax1}, the arrival time ($t_+$), and the total survey time ($T_\mathrm{sur}$, see Fig.~\ref{fig_1_y_max_from_delta_T}). We considered $T_\mathrm{sur}=5~\mathrm{years}$. The emission frequency enters in $y_\mathrm{min}$, which excises the region near the central caustic where wave optics is effective. Geometric optics is valid for $y>y_\mathrm{min}$.

Lens discovery rates are affected by the ability to resolve multiple images \citep{statistics}. Amplification has to push the signal above threshold, $A_{\pm}> \mathrm{SNR_\mathrm{th}^{\pm}}/\mathrm{SNR_{int}}$, which limits the source position to
\beq
\label{yMax1}
y_{\pm} < y_{\mathrm{max}}^{\pm} =  \min \left\{1, \left[  \left( \mathrm{SNR_\mathrm{th}^{\pm}}/\mathrm{SNR_{int}}\right)^2   \mp 1 \right]^{-1} \right\},
\eeq
according as the requirement is on the $+$ or the $-$ image, respectively. Multiple events are detectable if $y \le y_{\mathrm{max}}^-$, which goes to 1 only for $\mathrm{SNR_{int}} \gg \mathrm{SNR_\mathrm{th}^{-}}$.

Due to the finite duration of the survey, statistics of transient phenomena involve some missing events due to time delay \citep{statistics}. To observe both images, we require that $\Delta t< T_\mathrm{sur}-t_+$, which further constrains the source position. In Fig.~\ref{fig_1_y_max_from_delta_T}, we plot $y_\mathrm{max}$ due to the finite observation time as a function of $z_\mathrm{d}$. This constraint plays a role only for very massive lenses, which are very rare, or for very late arrival times.

\begin{figure}
        \scalebox{0.75}{\includegraphics{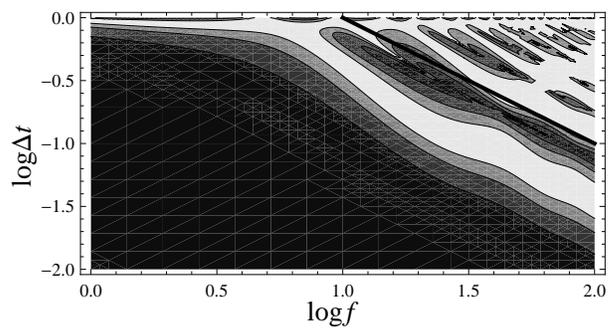}}
        \caption{
        {\bf  
        Relative error in the GW amplitude introduced by neglecting diffraction.
        } Contour values are 0.2, 0.1 and 0.05; darker areas correspond to larger errors. Time is in units of $\Delta t_z$. Above the full line, $\Delta t \times f >10$, geometric optics is valid.}
        \label{fig_2_diff_err}
\end{figure}

\begin{figure}
        \scalebox{0.75}{\includegraphics{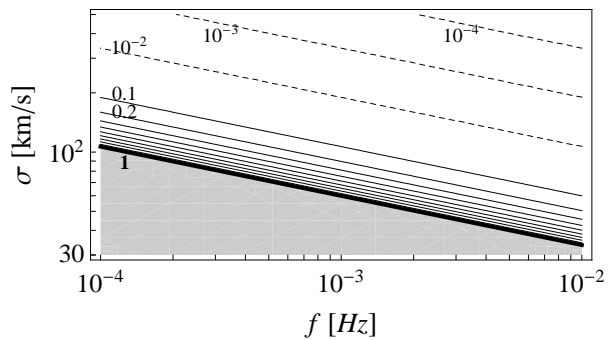}}
        \caption{Contour plot of $y_\mathrm{min}$ (such that $\Delta t \times f>10$) for $z_\mathrm{d}=2$ and $z_\mathrm{s}=6$. Full lines take values from 0.1 to 1 in steps of 0.1; dashed lines are for $y_\mathrm{min}=10^{-2},10^{-3}$ and $10^{-4}$.}
        \label{fig_y_min_from_diff}
\end{figure}

{\it Wave-optics.} Since we cannot observe multiple images when the interference pattern is pronounced, we have to discard configurations where wave effects are large  \citep{gw_diff}. Diffraction is negligible when the wavelength is much smaller than the gravitational radius of the lens, i.e., $\Delta t \gg f^{-1}$ \citep{gw_diff}. Geometric optics is valid if the coherence time of the signal is much smaller than the time delay between images, $ \Delta f \Delta t \gg 1$, where $\Delta f$ is either the bandwidth of the detector or the range in emission frequency. These conditions determine $y_\mathrm{min}$. The error in the estimated wave amplitude introduced by neglecting diffraction is displayed in Fig.~\ref{fig_2_diff_err}, where we exploited the sum representation of the diffraction integral \citep{gw_diff}.

Two features determine the effective frequency for MBHBs. First, most of the signal is emitted near the coalescence time. Second, LISA sensitivity drops for $f \gs f_\mathrm{ch}=3\times 10^{-3}~\mathrm{Hz}$. The effective frequency is then the minimum between $f$ at coalescence and $f_\mathrm{ch}$. Since the final frequency is much larger than the initial one, the two conditions give very similar constraints. Apart from a very small region near the central caustic, see Fig.~\ref{fig_y_min_from_diff}, wave effects are negligible. For a source $z_\mathrm{s}=6$ emitting at $f_\mathrm{ch}$ behind a massive lens ($\sigma \simeq150~\mathrm{km/s}$, $z_\mathrm{d}=2$), $y_\mathrm{min} \simeq 10^{-2}$.

\begin{figure}
    \scalebox{0.7}{\includegraphics{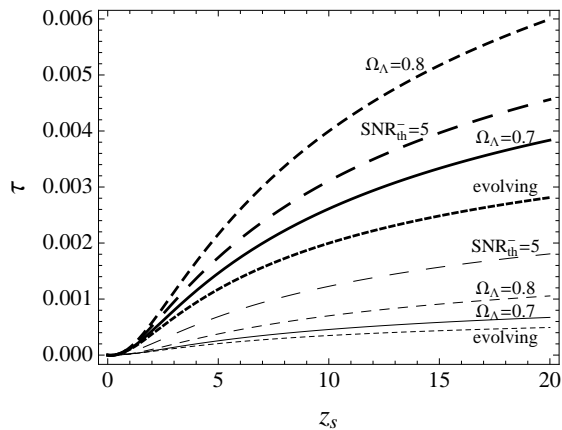}}
     \caption{$\tau$ for $f= f_\mathrm{ch}$ and $t_+=2~\mathrm{years}$. Thin and thick lines are for $\mathrm{SNR_{int}}=6$ or $20$, respectively. Full lines are for the standard case.
     }
        \label{fig_4_tau_SNR_6_20}
\end{figure}

\begin{figure}
       \scalebox{0.7}{\includegraphics{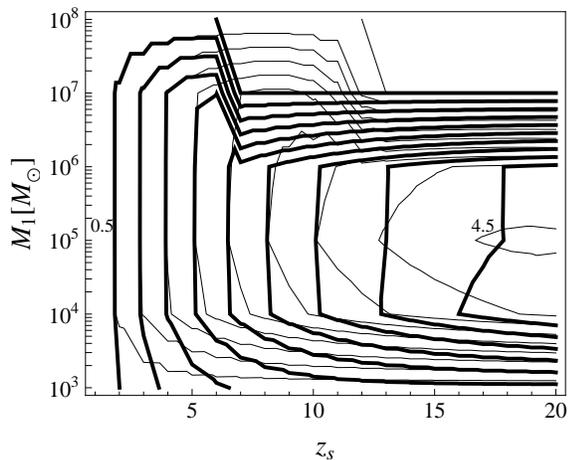}}
       \caption{$\tau$ contour plot for a MBHM with $f= f_\mathrm{ch}$ and $t_+=2~\mathrm{years}$. Thin and thick lines are for $M_2/M_1=0.1$ or $1$, respectively. Values go from 0 to $4.5\times10^{-3}$ in steps of $0.5\times10^{-3}$.}
        \label{fig_tau_zs_M1}
\end{figure}

{\it Optical depth.} 
The lensing probability for a source is the total optical depth $\tau$, obtained by integrating Eq.~(\ref{stat1}); it is plotted in Fig.~\ref{fig_4_tau_SNR_6_20} as a function of $z_\mathrm{s}$. In our standard case, $\Omega_\mathrm{M}=0.3$, the comoving $dn/d\sigma$ is constant and $\mathrm{SNR_\mathrm{th}^{\pm}}=8$ is the detection threshold. The more effective constraint on the cross section is the requirement of a loud enough second image. If we either lower $\mathrm{SNR_\mathrm{th}^-}$ or consider intrinsically brighter signals, $\tau$ dramatically increases. $\tau$ is very sensitive to the value of $\Lambda$ and goes down for a smaller number of lenses at high $z$. 

Probabilities are quite large for a range of BH binaries with masses ($10^4$-$10^7\msun$) and redshifts ($z\ls 15$) of astrophysical relevance in the LISA window, see Fig.~\ref{fig_tau_zs_M1}. $\tau$ is plotted in the standard case as a function of $z$ and the main BH mass, $M_1$, for two different mass ratios. The $\mathrm{SNR_{int}}$ of a given merger was computed according to \citep{lisa_sources}. Given $M_1$, the larger the mass ratio, the larger $\tau$. 

\begin{table*}
\centering
\begin{tabular}[c]{l c c | r@{$\,\pm\,$}l r@{$\,\pm\,$}l r@{$\,\pm\,$}l |  r@{$\,\pm\,$}l  r@{$\,\pm\,$}l  r@{$\,\pm\,$}l | r@{$\,\pm\,$}l r@{$\,\pm\,$}l r@{$\,\pm\,$}l  | r@{$\,\pm\,$}l r@{$\,\pm\,$}l r@{$\,\pm\,$}l}
        \hline
        \noalign{\smallskip}
        \multicolumn{3}{c}{scenario} &  \multicolumn{6}{c}{{\it Light} (90, $\bar{z}_\mathrm{s}=10.4\pm 4.1$)} & \multicolumn{6}{c}{{\it Heavy} (30, $\bar{z}_\mathrm{s}=5.8\pm2.4$)} & \multicolumn{6}{c}{{\it HybridI} (130, $\bar{z}_\mathrm{s}=7.2\pm3.8$)} & \multicolumn{6}{c}{{\it HybridII} (390, $\bar{z}_\mathrm{s}=6.0\pm3.4$)} \\
         \noalign{\smallskip}
        \hline
        $\Omega_\mathrm{M}$ &	$\frac{dn}{d\sigma}$ & 	$\mathrm{SNR_\mathrm{th}^-}$ & 	\multicolumn{2}{c}{$\langle N \rangle$} &	\multicolumn{2}{c}{$N\ge 1$} & 	\multicolumn{2}{c}{$N\ge2$} &	\multicolumn{2}{c}{$\langle N \rangle$} &	\multicolumn{2}{c}{$N\ge 1$} & 	\multicolumn{2}{c}{$N\ge2$}  & 	\multicolumn{2}{c}{$\langle N \rangle$} &	\multicolumn{2}{c}{$N\ge 1$} & 	\multicolumn{2}{c}{$N\ge2$}  & 	\multicolumn{2}{c}{$\langle N \rangle$} &	\multicolumn{2}{c}{$N\ge 1$} & 	\multicolumn{2}{c}{$N\ge2$} \\
        \noalign{\smallskip}
        \hline
        0.3 &		co. &		8 & 	.38 & .03 &	32 & 2 & 	5.8 & .7	&	.24 & .02 &	21 & 1& 	2.4 & .3  &	.55 & .04 &	42 & 2& 	10.6 & 1.2	&	2.83 & .09 &	94.1 & .5& 	77.4 & 1.5\\
         \hline
         0.2 &	co. &		8 & 	.58 & .04 &	44 & 2 & 	11 & 1 	&	.35 & .03 &	29 & 2& 	4.8 & .6  &	.81 & .06 &	55 & 3& 	19 & 2	&	4.17 & .14 &	98.4 & .2& 	92.0 & .9\\
         \hline
         0.3 &	ev. &		8 & 	.30 & .02 &	26 & 1 & 	3.7 & .4 	&	.19 & .01 &	17 & 1& 	1.5 & .2  &	.44 & .03 &	35 & 2& 	7.2 & .9	&	2.24 & .07 &	89.4 & .7& 	65.7 & 1.7\\
         \hline
         0.3 &	co. &		5 & 	.52 & .03 &	41 & 2 & 	10 & 1 	&	.24 & .02 &	21 & 1& 	2.5 & .3  &	.68 & .05 &	49 & 2& 	14.7 & 1.6	&	3.06 & .10 &	95.3 & .5& 	80.9 & 1.4\\
         \hline
         \hline
\end{tabular}
\caption{
Lensing probabilities assuming $\mathrm{SNR_\mathrm{th}^{+}}\ge 8$. $\langle N \rangle$ and chances (in \%) to observe one or more lensing events, $N\ge 1,2$, for different hypotheses (name, number of events per year, typical source redshift and its dispersion). Biweight estimators are reported. 
}
\label{tab_tau}
\end{table*}

\begin{figure}
      \scalebox{0.75}{\includegraphics{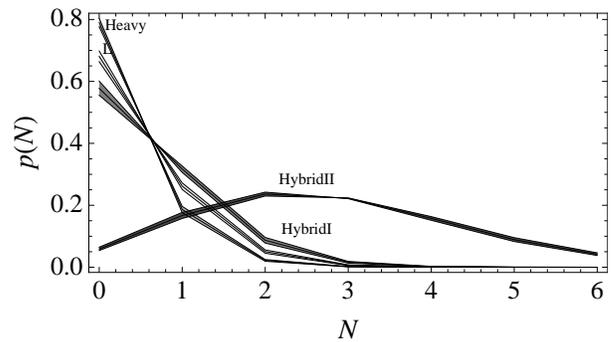}}
        \caption{Probability of detecting $N$ lensing events in the standard case; 
        thick curves include 68\% confidence regions; ``L'' = {\it Light}.
        }
        \label{fig_pdf_number}
\end{figure}

{\it LISA sources.}
To estimate the lensing probabilities we need plausible merger rates. We adopt MBH assembly histories created via dedicated Monte Carlo merger tree simulations framed in the hierarchical paradigm \citep{mergers}. Either first seed BHs are light, $M\gtrsim100\msun$, being the remnant of the first POPIII star explosions \citep{mr01} or already quite heavy ($M\gtrsim10^4\msun$) seeds form by direct collapse of proto-galactic discs \citep{heavy}, or through efficient accretion onto a ``quasi star'' forming in dense disks prone to bars-within-bars instabilities \citep{bvr06}. The first two models we analyzed ({\it Light} and {\it Heavy} models, LE and SE in \citep{aru09}, respectively) have been adopted as a benchmark for model comparison by the LISA parameter estimation taskforce. Furthermore, we considered a pair of ``hybrid'' models where the two formation processes coexist and the seed mass spectrum spans the range $10^2$-$10^5\msun$ \citep{vb10}. The two hybrid models have somewhat different formation efficiencies, which are set by the maximum spin parameter $\lambda_\mathrm{Thr}$ of the halo that allows an efficient inflow of gas in the nuclear region. We considered a conservative version ($\lambda_\mathrm{Thr}=0.01$, {\it HybridI}) and a more optimistic one ($\lambda_\mathrm{Thr}=0.02$, {\it HybridII}). 

All models have been extensively tested against observational constraints, such as the present day mass density of nuclear MBHs, the optical and X-ray luminosity functions of quasars and the unresolved X-ray background \citep{mergers}. {\it Light}, {\it Heavy} and {\it HybridI} models are on the conservative side, while {\it HybridII} is more optimistic. LISA source redshifts are much higher than usual, see Table~\ref{tab_tau}, and allow to investigate the Universe to very large distances. By comparison, radio-sources in the CLASS survey had $\bar{z}_\mathrm{s}=1.4\pm0.7$ \citep{statistics}.

{\it Expected events.} 
The mean number of lenses expected to be observed, $\langle N \rangle$, is given by the sum of the optical depths of all the sources. We simulated the intrinsic properties instead of the observed distribution of amplitudes so that we did not have to correct for any magnification bias \citep{statistics}. Results are listed in Table~\ref{tab_tau}. The Poissonian probability for multiple events is plotted in Fig.~\ref{fig_pdf_number}.  Chances are fairly large for all the build-up scenarios. Wave effects are negligible: in the geometric optics limit ($y_\mathrm{min}=0$), $\tau_\mathrm{tot}$ increases by $\ls 0.01\%$.

The bulk of the total cross section is due to high $\mathrm{SNR_{int}}$ events. In the {\it Light} scenario, the fraction of events ($\sim 54\%$) intrinsically below threshold contributes only $\ls 8\%$ of $\tau_\mathrm{tot}$. Due to the many events with very high $\mathrm{SNR_{int}}$, results for the {\it Heavy} case are not affected at all by lowering $\mathrm{SNR_\mathrm{th}^-}$, whereas probabilities increase significantly in the {\it Light} and {\it HybridI} cases. LISA should detect independently two above threshold signals in the same position in order to claim lensing. Any dedicated pipeline for detection of double peaked events might strong enhance lensing detection.

A larger, but still compatible with observations, $\Lambda$ strongly increases the probabilities. New merger trees should be generated, but to stress the role played by a larger $\Lambda$ we focused on the variation of $\tau$. Reducing the number of lenses at high $z$ has an opposite effect but chances are still sizable. 

Lensing amplification can make some intrinsically too faint mergers luminous enough to be detected. If we are not interested in the second image, $y_\mathrm{max}$ can be larger than one and is determined only by the flux condition on the $+$ image. In the {\it Light} ({\it HybridI}) model, nearly $3(2)\%$ of the otherwise undetected events, $\gs 1$ event per year, may become loud enough to be seen. In the {\it HybridII} scenario, $8\pm3$ additional events per 5 years are expected.

We provided one of the first attempts at formalizing lensing statistics in presence of wave-effects. Our assumptions were conservative: a 5~years mission; high detection thresholds; not so optimistic formation scenarios; no account of lensing by late-type galaxies, whose distribution is poorly known but contribute $ \ls 30\%$ of $\tau_\mathrm{tot}$ \citep{statistics}. Probabilities to observe multiple events were anyway sizable, from $\ls 20$ to $\ls 100\%$.  Lensing statistics depend on the cosmological parameters, suggesting new potential tools for cosmography with LISA. In case of optical identification of lenses, additional measurements of time-delay might help to constrain $H_0$ and dark energy. $\tau$ is very sensitive to $d n/d \sigma$ too, which might help to understand evolution of early-type galaxies up to high redshifts.

The main unsureness in our forecast is due to the uncertain build-up process. Tight predictions need knowledge of galaxy evolution at high $z$, one of the main LISA goals. To establish that lensing probabilities for LISA might be significant and to point to the relative rewards, it was enough to consider a broad range of formation scenarios without trying to maximize $\tau_\mathrm{tot}$. Hopefully, LISA lensing might help to constrain the formation history.

Apart from astronomical applications, the detection of even a single GW lensing event would provide unique information on competing theories of gravity by allowing to test velocity and propagation of gravity. Up to now, GWs are still missing on an experimental ground. Their mere detection by LISA would be a success. Serendipitous GW lensing might be an extremely beneficial bonus.


\end{document}